\documentclass[11pt]{article}
\usepackage[sc]{mathpazo}
\usepackage{fullpage}
\usepackage[authoryear,sectionbib,sort]{natbib}
\usepackage{graphicx}
\usepackage{adjustbox}
\usepackage[boxruled,linesnumbered]{algorithm2e}
\usepackage{multirow}
\usepackage{amsmath}
\usepackage[utf8]{inputenc}
\usepackage{lineno}
\usepackage{titlesec}
\usepackage{graphicx}
\usepackage{subcaption}
\usepackage{booktabs,multirow}
\usepackage{here}
\usepackage{url}

\newcommand{\mat}[1]{\mbox{\boldmath{$#1$}}} 

\usepackage{listings}
\usepackage{xcolor}

\lstdefinestyle{mystyle}{
	backgroundcolor=\color{backcolour},   
	commentstyle=\color{codegreen},
	keywordstyle=\color{magenta},
	numberstyle=\tiny\color{codegray},
	stringstyle=\color{codepurple},
	basicstyle=\ttfamily\footnotesize,
	breakatwhitespace=false,         
	breaklines=true,                 
	captionpos=b,                    
	keepspaces=true,                 
	numbers=left,                    
	numbersep=5pt,                  
	showspaces=false,                
	showstringspaces=false,
	showtabs=false,                  
	tabsize=2
}

\definecolor{codegreen}{rgb}{0,0.6,0}
\definecolor{codegray}{rgb}{0.5,0.5,0.5}
\definecolor{codepurple}{rgb}{0.58,0,0.82}
\definecolor{backcolour}{rgb}{0.95,0.95,0.92}

\titleformat{\section}[block]{\Large\bfseries\filcenter}{\thesection}{1em}{}
\titleformat{\subsection}[block]{\Large\itshape\filcenter}{\thesubsection}{1em}{}
\titleformat{\subsubsection}[block]{\large\itshape}{\thesubsubsection}{1em}{}
\titleformat{\paragraph}[runin]{\itshape}{\theparagraph}{1em}{}[. ]

\title{Understanding Social Immunity in Ants: A Markovian Approach to Collective Cleaning Strategies}

\author{Isabella Bueno$^{1\ast}$ \and 
Gabriel R. Palma$^{2, 3}$ \and
Idemauro A. R. Lara$^{4}$  \and
Rafael A. Moral$^{2, 3}$\and
Italo Delalibera Jr$^{1}$ \and
Wesley A. C. Godoy$^{1}$}

\date{}

\begin{document}
\maketitle
\noindent{} 1. Department of Entomology and Acarology, University of S\~{a}o Paulo, Piracicaba, S\~{a}o Paulo, Brazil;

\noindent{} 2. Hamilton Institute, Maynooth University, Maynooth, Ireland;

\noindent{} 3. Department of Mathematics and Statistics, Maynooth University, Maynooth, Ireland;

\noindent{} 4. Department of Mathematics and Statistics, University of S\~{a}o Paulo, Piracicaba, S\~{a}o Paulo, Brazil;

\noindent{} $\ast$ Corresponding author; e-mail: ibueno@usp.br

\bigskip


\bigskip


\bigskip

\textit{Manuscript type}: Research paper. 

\bigskip

\noindent{\footnotesize Prepared using the suggested \LaTeX{} template for \textit{Am.\ Nat.}}

\newpage{}

\section*{Abstract}
Understanding social immunity mechanisms in ant colonies remains crucial to comprehending the evolution of defense strategies in eusocial organisms. This study assumes the absence of the role of memory in the ants' defense strategy, considering that they can make a new exploration of collective cleaning behaviors. We investigate how worker interactions and previous behaviors influence the evolution of social immunity strategies in response to the presence of pathogens and vulnerable colony members. In this context, with the application of Markov transition models it was possible to describe changes in cleaning behavior over time influenced by the presence of vulnerable members and treatment with an entomopathogenic fungus. We found a significant effect on ant behavior when exposed to \textit{Metarhizium anisopliae} when a fungus garden fragment and one larva were present. Our findings confirm a link between prophylactic cleaning and the presence of vulnerable members. Remarkably, distinct behaviors and transition times vary among treatments, revealing workers’ adaptability to threats. Allogrooming shows adaptive changes when exposed to pathogens, potentially affecting pathogen transmission. In addition, our study elucidates the intricate interaction between internal and external factors shaping worker behavior. The influence of environmental context on decision-making principles emerges, emphasizing the importance of both intrinsic colony organization and external threats. By using a Markov model to understand ant hygiene behavior, we offer insights into social immunity mechanisms in eusocial organisms. Deciphering collective cleaning strategies can aid in understanding disease dynamics and decision-making processes in complex societies.

\newpage{}

\section{Introduction} 
Leafcutter ants are part of the group of fungus-farming ants~\cite{wetterer1998,ward2014} that exclusively utilize leaves cut by workers to cultivate their symbiotic fungus~\cite{weber1966}. Cultivating fungi for sustenance represents an ecological relationship that originated among a particular group of ants in South America approximately $53.6$ to $66.7$ million years ago~\cite{jesovnik2016}. These ants, recognized as ecologically dominant herbivores, form expansive "superorganisms" comprising millions of individuals ~\cite{branstetter2017dry}.

\textit{Atta} Fabricius is one of the three extant genera of leafcutter ants situated within the subtribe Attina (Hymenoptera: Formicidae: Myrmicinae) ~\cite{sosacalvo2017attina}. Due to their communal lifestyle, this group forms densely populated colonies of closely related individuals, rendering them susceptible to parasites ~\cite{schmidhempel1998,boomsma2005}. Various parasites, notably entomopathogenic fungi, exhibit close associations with social insects ~\cite{schmidhempel1998}. In the context of leafcutter ants, defending against pathogens is a critical factor for sustaining their symbiotic fungus.

The mutualistic relationship of attine ants with the fungus \textit{Leucoagaricus gongylophorus} ~\cite{weber1966,mehdiabadiandschultz2010} increases the complexity of interactions between this group of ants and microorganisms. In addition to safeguarding their species from microbial threats, worker ants must also defend the fungus garden. This mutualism is mandatory for both the ants and the fungus. The ants rely on the fungus for shelter, serving as the exclusive food source for their larvae and providing partial nourishment for the workers ~\cite{QuinlanandCherrett, bassandcherrett1995}. Simultaneously, the fungus depends on the ants for care and nutrient supply ~\cite{theants, holldoblerwilson2011}.

This pronounced interdependence between the fungal garden and the ants, coupled with the clonal nature of the gardens, causes the fungus to rely heavily on the workers for defense against microorganisms. Among the defense strategies ants employ is promoting genetic diversity through the horizontal transfer of fungal cultivars between attine species ~\cite{mueller2018phylogenetic}. Additionally, they manage the garden by isolating it in chambers to prevent contamination, monitoring, weeding, removing pathogens, and applying antibiotics to suppress invasive microorganisms, among other techniques ~\cite{mueller2005evolution}.

To protect themselves and vital colony members like the queen, offspring, and the fungus garden, worker ants possess an array of physiological and behavioral components to combat pathogens. A primary physiological defense involves the production of antimicrobial compounds by the metapleural gland  ~\cite{fernandez2006active, fernandez2015functional}. Workers utilize cleaning behaviors to selectively disperse the antimicrobial compounds produced by the metapleural glands throughout the colony, especially when facing pathogenic challenges ~\cite{fernandez2006active}. Through specific leg movements, workers clean the openings of the metapleural glands, collect the secretions, and transfer them to their mouthparts, allowing the antibiotic substances to be distributed in the colony’s vicinity ~\cite{fernandez2003revision, fernandez2006active}.

Cleaning behaviors, as observed in worker ants, constitute a pivotal strategy for managing pathogenic microorganisms, assisting in addressing infections impacting the workers and the fungal garden ~\cite{currie2001weeding, jaccoud1999epizootiology}. These behaviors intensify particularly when confronted with a fungal infection ~\cite{currie2001weeding, jaccoud1999epizootiology,nilsson2018visual,morelos2010ant}. Furthermore, in response to the presence of pathogens, not only do ants engage in reactive cleaning behaviors but also exhibit proactive actions aimed at protecting vulnerable colony members (e.g., offspring) and the fungal garden, serving to prevent contaminations  ~\cite{morelos2010ant}.

Additionally, these social insects deploy collective defenses alongside individual physiological mechanisms to combat pathogens. Notably, mutual cleaning behaviors such as allogrooming exemplify the effectiveness of communal pathogen removal and the promotion of collective immunity  ~\cite{cremer2007social, cremer2018social}. Termed ”social immunity,” this suite of defenses significantly reduces group mortality ~\cite{cremer2007social}.

The extensively studied phenomenon of social immunity ~\cite{cremer2009analogies, ugelvig2010rapid, cremer2018social, meunier2015social, pull2018destructive, stroeymeyt2018social} coupled with individual defense strategies ~\cite{schluns2009molecular, konrad2012social, tragust2013ants, nilsson2018visual} reveals the comprehensive exploration in this field. Experiments involving the recording of time-related response categories, such as insect behavior, is a common approach to verifying treatment effects and forecasting behavioral transitions. Analyzing longitudinal categorical data necessitates employing models that accommodate the correlation structure within dependent data. In this context, multi-state models are a particular class of models that allows for describing the changes of response categories over time, conditional to the effects of time and treatments. These models are based on stochastic processes and help predict behavioral transitions ~\cite{meira2009, lara2020}. 

This study aimed to apply the multi-state Markov model to explain the decision-making process of cleaning behaviors in leaf-cutter ant workers under different experimental conditions. These models are based on continuous stochastic process \cite{ross1995} and they are instrumental in analyzing behavior over time  \cite{lara2017}. We carried out an experiment to measure the cleaning behavioral response of ants when exposed to the entomopathogenic fungus \textit{Metarhizium anisopliae}, both in the presence and absence of vulnerable colony elements, such as brood and the fungus garden.

\section{Methods}
\subsection{Species studied}

The \textit{Atta sexdens} ants used in the study were obtained from colonies collected by excavating the soil in a teak plantation (\textit{Tectona grandis}) in Anhembi, State of São Paulo, Brazil (22°40’S 48°10’W). After collection, colonies were kept in the laboratory under controlled temperature and light conditions. Five colonies of approximately two years of age were used to supply the workers.

\subsection{Fungal suspension}
The \textit{Metarhizium anisopliae} fungus, isolate ESALQ E9, is deposited in the Entomopathogen Collection ”Prof. Sérgio Batista Alves” from the Luiz de Queiroz College of Agriculture – University of S\~ao Paulo (ESALQ-USP), Brazil. Seven days before the assay, the Petri dishes were replicated from samples kept in a $-80$ºC freezer in cryotubes. The fungus was cultivated in PDA medium (potato dextrose agar, KASVI®). The Petri dishes (8 x 1,5cm) were kept in BOD chambers with a controlled temperature ($25$ºC) and a photoperiod of $12$ hours for seven days. Suspensions were prepared using conidia scraped from the Petri dishes and added to 10ml of Tween $80$º surfactant. The concentration used was $10^8$ conidia/mL.

\subsection{Behavioral experiment}
To study the grooming behaviors of leaf-cutting ants under different conditions, an experiment was set up using a randomized complete block design, with five blocks. Each block corresponded to the particular colony from which workers were removed. Each block was repeated twice, totalling 10 replicates of four treatments: T1 = five workers exposed to \textit{Metarhizium anisopliae}; T2 = five workers exposed to Tween $0.05\%$; T3 = five workers exposed to \textit{Metarhizium anisopliae} with the presence of a fungus garden fragment and one larva; and T4 = five workers exposed to Tween $0.05\%$ with the presence of a fungus garden fragment and one larva. Each experimental unit consisted of a $9$ cm diameter round Petri dish lined with a plaster.

The worker ants used in the experiment were medium-sized, and painted in different colors with an insect marking pen, to allow for distinguishing each one. The painted workers were added to Petri dishes $24$ hours before the experiment and kept for acclimation. The experiment was carried out under controlled conditions of temperature ($25$ºC) and humidity ($70\%$). The workers were anaesthetized with CO2 before the application of the treatments. Exposure to treatments was performed by placing workers on a $10$-minute walk on filter paper with 1 ml of fungal suspension or Tween.

Each experimental unit was filmed for 30 minutes after the ants walked in the Petri dish. The captured images were taken with a Sony Handycam camcorder and indirect artificial lighting. The camera was placed perpendicularly above the Petri dish at an angle of $90$º. The grooming behaviors (allo and self-grooming) were recorded and quantified using the Behavioral Observation Research Interactive Software (Boris) program, version 7.13.8 ~\cite{friard2016boris}. The observed behaviors were based on the descriptions presented by ~\cite{nilsson2018visual}: allogrooming was considered when one or more ants were close to each other, with physical contact. The `cleaner' ant moved slightly to cover the receiving ant's body area; self-grooming was recorded when the worker cleaned its antenna,  the antenna cleaners, and later the ant cleaned the legs, taking the legs to the mouthparts and removing particles and pathogens with the glossa.

Other behaviors were rare or sporadic; therefore, we classified them as "other" when a particular worker was not engaged in self-grooming or allogrooming, thus consisting a categorical response variable with three categories. Most of the ``other'' behaviors involved walking and staying at the edges of the Petri dish. After the end of the recording sessions, the workers were kept in their respective Petri dishes inside humid chambers ($25$ºC and $90\%$), and the mortality was counted every two days. The dead workers were removed, washed in $70$º alcohol, and later in distilled water, and kept in a humid chamber for 12 days for spore growth and confirmation of death by the applied entomopathogenic fungus. 

\subsection{Statistical analysis}
\begin{figure}
    \centering
    \includegraphics[width=12cm]{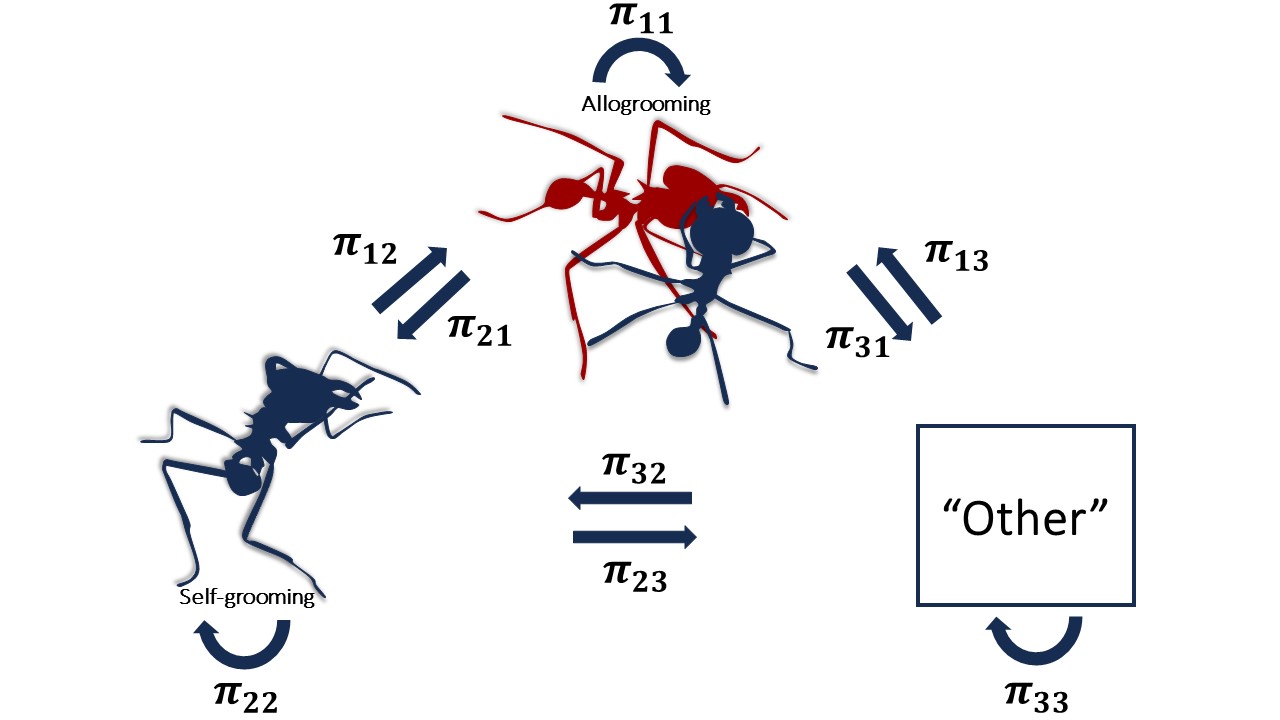}
    \caption{Illustrative diagram of the transition probability matrix representing the probability $\pi_{ab}$ of changing from behavior $a$ to behavior $b$.}
    \label{Esquema}
\end{figure}

We considered multi-state models \cite{meira2009} to jointly model the behavior and time associated with each ant action. Our objective with this modeling approach was to describe changes in response categories (behavior transitions), time spent in each state, and the effects of treatments on the process. These models are based on continuous-time stochastic processes \cite{ross1995}. The process is represented by the
 set $\{Y(t), t \in \tau \}$, where
$Y(t) \in S=\{1, 2, 3\}$, that represents the categories of ant behavior~(1: allogrooming, 2: self-grooming and antenna cleaning, or 3: other) and $\tau=[0, 30)$ the observation time (Figure~\ref{Esquema}). To improve model convergence and favor parsimony, we assumed that the behavior transition probabilities are given by the Markovian property:

\begin{eqnarray}\label{mp}
			\pi_{ab}(t)=P(Y_{(t+s)}=b \mid Y_{(s)}=a), \hspace{0.2cm} \forall \hspace{0.2cm}  s< t \in  \tau \hspace{0.2cm}  \mbox{and} \hspace{0.2cm}  a,b \in S. 
				\end{eqnarray}
The transition intensities are such that
		\[  \theta_{ab }=   \left\{ \begin{array}{ll}
			-\theta_{a}  & \mbox{if} \hspace{0.3cm} k = k'    \\
			\theta_{a} \pi_{ab}(t) & \mbox{if} \hspace{0.3cm}  k \neq k',   \\
		\end{array} \right. \]  \\
where $k$ is the dimension of the square intensity matrix $\mathbf{Q}$. For our experiment, $k=3$, and therefore we have
\begin{eqnarray} \nonumber
\mathbf{Q}=\left(\begin{array}{ccc}
		-(\theta_{1}+\theta_{2}) & \theta_{1}&\theta_{2} \\
		\theta_{3} & 	-(\theta_{3}+\theta_{4})  &\theta_{4} \\
		\theta_{5}& \theta_{6} &	-(\theta_{5}+\theta_{6})  \\
	\end{array}\right).
\end{eqnarray}
The transition intensities are the parameters associated with time, such that $\displaystyle{-\frac{1}{\theta_{a}}}$ is the mean time spent in behavior $a$.
To consider the effect of experimental design (blocks and treatments), the multi-state model considered was the Cox (proportional hazards model), given by:
\begin{eqnarray}\label{cox_model}
	\theta_{ab}(\mat{x}) = \theta_{ab}^{(0)}\exp\{\mat{\beta}_{a}^{\top}\mat{x}\}
\end{eqnarray}
where $\theta_{ab}^{(0)}$ is the initial estimate for the transition intensities matrix, $\mat{x}$ is the vector that describes the experimental design and $\mat{\beta}_{a}$ is the associated vector of parameters, that depends on $a \in S$.	

The Cox model assumption was tested by the hazard proportionality test proposed by~\cite{grambsch1994}. Also, the parameters are estimated through iterative processes, where the intensity matrix is updated at each step. Once convergence is obtained, the probabilities given by (\ref{mp})  are estimated from (\ref{cox_model}), yielding the estimated transition probability matrix
$$\mat{\hat{P}}(s,t)=\exp[(t-s)\mat{\hat{Q}}].$$ These models were fit and assumptions were tested using the \texttt{msm} ~\cite{jackson2011} and \texttt{survival} packages~\cite{survival2023}, available for \texttt{R} software \cite{R2022}. To allow for full reproducibility of our results, the data and R scripts are available at \url{https://github.com/GabrielRPalma/UnderstandingComplexBehavior}.

\section{Results}
The most common behaviors observed in were self-grooming, allogrooming, and walking. Other behaviors such as offspring care, fungus grooming and weeding were rare. As other behaviors were rare, we delimited the category 'others' for any behavior not defined as self-grooming or allogrooming. A total of 3897.2 minutes of behavior were recorded, with $5.76\%$ being allogrooming, $14.61\%$ self-grooming, and $79.63\%$ classified as 'others'.  With respect to changes in the behavior of ants, we observed 2,858 first-order transitions: 214 from allogrooming, 1,216 from self-grooming, and 1,428 from `other' behaviors. The distribution of time spent in each behavior was skewed, especially for `other' behaviors. 

\begin{figure}[!ht]
    \centering
    \includegraphics[width=12cm]{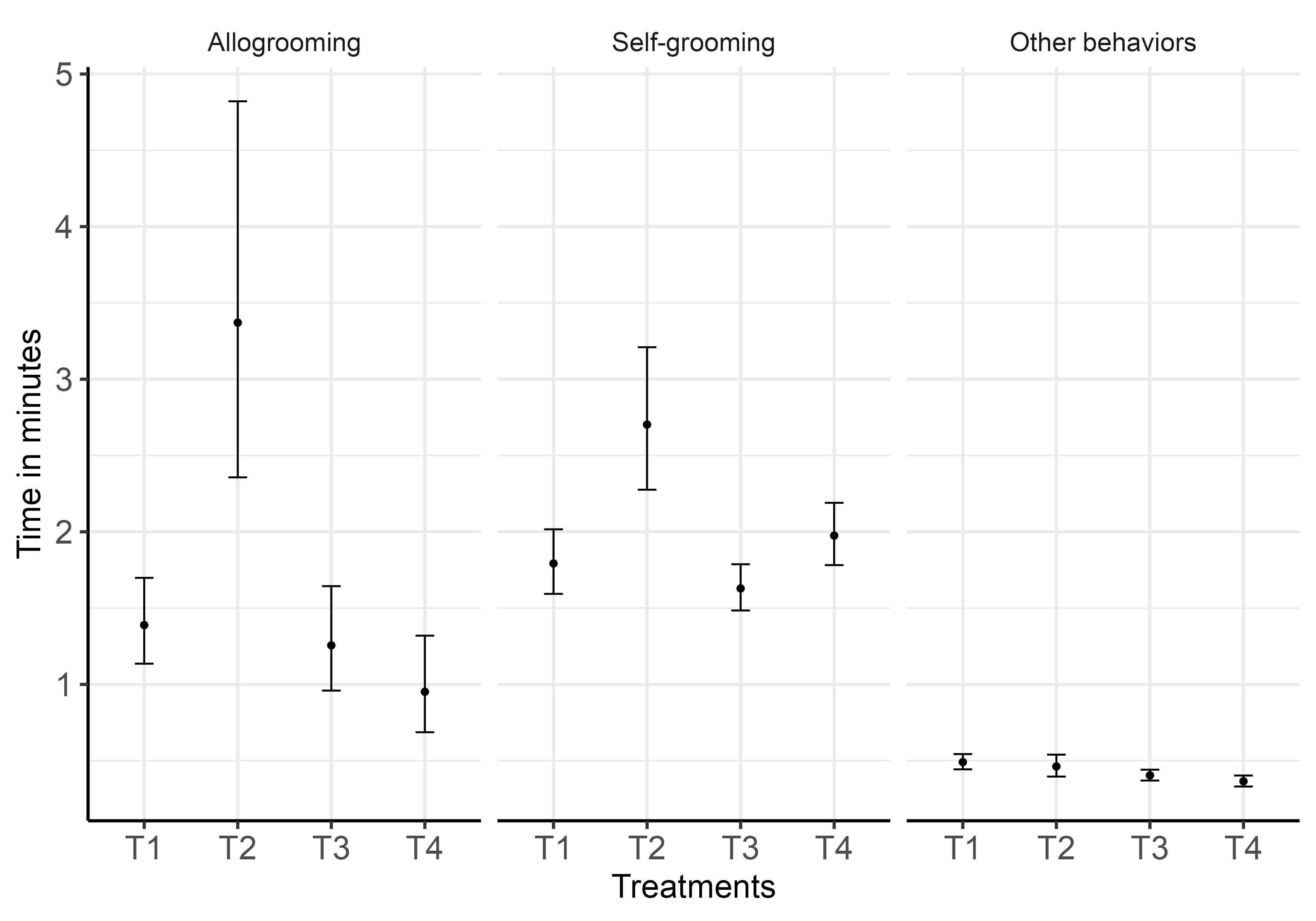}
    \caption{{\small  Time point estimates with their respective $95\%$ confidence interval based on the multi-state model for each behavior and treatments (T1 = 5 workers exposed to \textit{Metarhizium anisopliae}; T2 = 5 workers exposed to Tween $5\%$;
T3 = 5 workers exposed to \textit{Metarhizium anisopliae} with the presence of a fungus garden fragment and one larva per plate; and T4 = 5 workers exposed to Tween $5\%$ with the presence of a fungus garden fragment and one larva per plate).}}
    \label{FiguresBarras}
\end{figure}

We fail to reject the null hypothesis that hazard rates are not proportional (LR = $3.43$, df = $3$, $p = 0.330$). Therefore, the assumption of the Cox model is adequate. The time point estimates and $95\%$ confidence interval constructed using the fitted multi-state model that insects have spent in each behavior given the treatment effect are shown in Figure~\ref{FiguresBarras}.



\begin{figure}
    \centering
    \includegraphics[width=12cm]{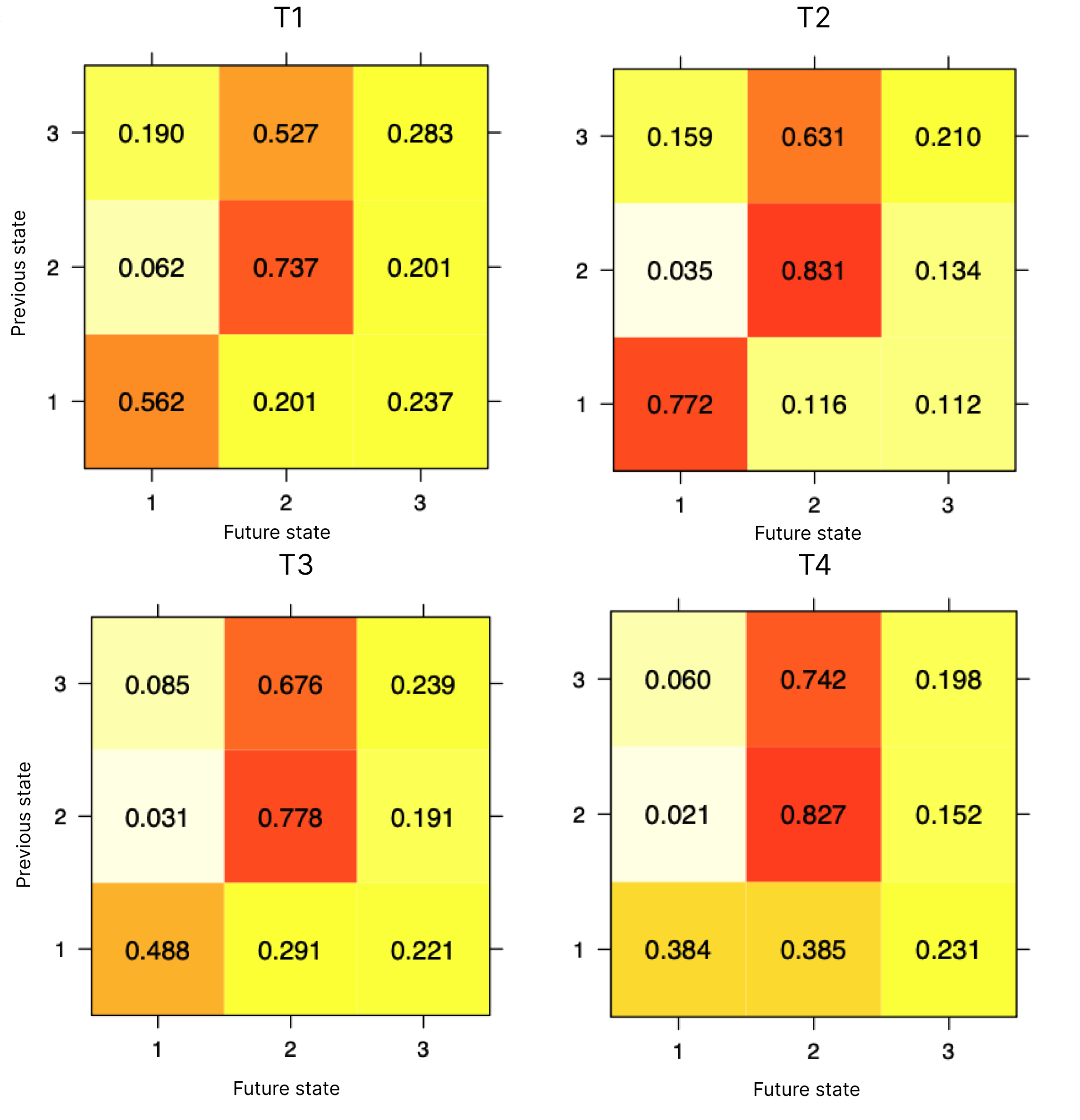}
    \caption{Estimated transition probability matrices for each treatment of the experiment according to the fitted multi-state model for each of three types of ant behavior: 1 -- Allogrooming, 2 -- self-grooming and antenna cleaning or 3 -- other behaviors. T1 = \textit{Metarhizium anisopliae}; T2 = Tween $0.05\%$; T3 = \textit{Metarhizium anisopliae} with the presence of a fungus garden fragment and one larva per plate; and T4 = Tween $0.05\%$ with the presence of a fungus garden fragment and one larva per plate. The color gradient varies with the increasing probability of transitioning from one behavior to another. Warmer colors correspond to a higher transition probability, while lighter colors correspond to lower transition probabilities.}
    \label{ProbabilityMatrices}
\end{figure}

We found a significant effect of treatments (LR = $97.52$, df = $18$, $p < 0.01$) according to the fitted multi-state model. In general, the estimated transition matrix shows how the choice of behavior is influenced by the behavior performed by the same individual one time step before and how the treatment influences this transition probability. In T1, composed only of workers exposed to the entomopathogenic fungus (Fig.~\ref{ProbabilityMatrices}), the highest transition probability found was that workers engaged in self-grooming would remain in the same behavior, followed by staying in the behavior of allogrooming. The lowest transition probability in this treatment was that workers engaged in self-grooming would shift to allogrooming behavior. In Treatment 2, workers were exposed only to the control treatment (Fig.~\ref{ProbabilityMatrices}), the highest transition probability found was that workers engaged in self-grooming would remain in the same behavior. The lowest transition probability in this treatment was that workers engaged in self-grooming would stay on the same behavior.

Treatments T3 and T4, with workers exposed to the entomopathogenic fungus and the control treatment, respectively, in addition to the presence of the fungus garden and brood, showed similar results regarding the transition probability between behaviors. The highest probabilities of behavioral transition from one-time step before to the next were that workers engaged in self-grooming would remain in that behavior. Next, the highest transition probability found was that workers engaged in the behavior categorized as ”other” would transition to self-grooming behavior. The lowest transition probability in this treatment was that workers engaged in self-grooming would remain in the same behavior.

\section{Discussion}
Although memory is expected to play a significant role in the social immunity mechanisms of ant colonies ~\cite{goes2020leaf}, this study considers ants as "Markovian organisms", meaning without memory ~\cite{lumsden1982social}. Consequently, it is possible to model the cleaning behavior of ants to predict which collective cleaning strategies may have evolved to achieve optimal social immunity, considering solely the influence of worker interactions among themselves and with the environment, as well as the occurrence of behaviors at a previous time step.

Our results confirm the assumption that prophylactic cleaning occurs in the presence of the most vulnerable members of the colony and increases as some pathogen is encountered. Furthermore, our model demonstrated the differences in the transition time of each behavior in different treatments. For both allogrooming and self-grooming behaviors, the longest transition time was found in treatment T2, indicating that when workers are not in the presence of the fungus garden and brood or entomopathogenic fungus, they are more inclined to remain in the same behavior. Concerning self-grooming, the shortest transition time was found in treatment T3, where the fungus garden/brood and entomopathogenic fungus were present, meaning each worker spent less time engaged in a specific task before switching to another. Regarding allogrooming, the two shortest transition times were in treatments T3 and T4, where the fungus garden and brood were present, with and without the pathogen. We can infer that when the threat or vulnerable component is present, the transition time to another activity decreases, indicating that each worker needs to alternate behavior, possibly to increase cleaning efficiency.

Despite a few occurrences of allogrooming, this is an important behavior that allows workers to clean each other in hard-to-reach places. Despite its importance, allogrooming may have a disease-spreading effect. Workers exposed to the pathogen tend to increase self-grooming while decreasing allogrooming adaptively to contain the pathogen spread ~\cite{Theis2015}. We found that in treatment T2, where there was no fungus garden or pathogen, workers spent more time continuously performing allogrooming before moving to another behavior. We can infer that in the absence of vulnerability or threat, workers tend to remain in the same cleaning behavior for allogrooming and self-grooming. As it involves two or more workers, the allogrooming rate and duration can alter the colony's social contact networks. Apart from the intrinsic organization of social insects to avoid harmful microorganisms  ~\cite{hart2001task}, behavioral modulation in the face of a pathogen affects interaction dynamics, impacting the risk of pathogen exposure for other workers and disease transmission  ~\cite{Theis2015}.  Colonies exposed to pathogens have their social contact network altered to minimize disease transmission, including protection for valuable colony components such as the queen ~\cite{baracchi2014socio, stroeymeyt2018social}.

The transition probability matrix showed significant differences across the four treatments. This matrix demonstrates the likelihood of the current behavior being influenced by the behavior one time step before. Additionally, these probabilities are influenced by the treatment. In the two treatments where vulnerable members were present (T3 and T4), the lowest probabilities were found for a worker engaged in allogrooming to remain in that state, as well as the lowest probabilities for workers engaged in the other two behaviors to transition to allogrooming. This lower transition probability to allogrooming in the presence of vulnerable members may be directly related to the potential role of allogrooming in disease dissemination ~\cite{Theis2015}, as mentioned earlier.

Despite the matrix presenting significant differences across all treatments, a pattern can be observed. In all four treatments, the highest probability found was for individuals engaged in self-grooming to remain in self-grooming, while the lowest probability found was for individuals engaged in self-grooming to transition to allogrooming. Moreover, the transition probabilities were generally considerably higher for persisting in a state than for switching states, even across different treatments. This might indicate fixed mechanistic principles governing individual worker decision-making, albeit with varying intensities depending on the environmental context.

Although cleaning behaviors are partially individual, the decision-making process that leads to social immunity involves several factors. These factors can be both external, such as the presence of microorganisms ~\cite{jaccoud1999epizootiology, currie2001weeding}, and internal, such as the number of present workers ~\cite{hughes2002trade}, presence of the queen ~\cite{keiser2018queen}, presence of other vulnerable colony members ~\cite{morelos2010ant}, among others. The presence of valuable colony members increases the disease cost; therefore, an optimization of defense strategy is expected. In this regard, an ant colony follows a system of stigmergic decision-making ~\cite{friedman2021active}. Essentially, stigmergy describes a mechanism of indirect communication among individuals, where the actions of one organism alter the environment and influence the future behavior of other individuals in the same shared environment. Thus, stigmergy involves decentralized coordination, allowing for the coordination of complex activities among eusocial organisms ~\cite{heylighen2016stigmergy}.

Insect societies can serve as models for studying disease spread in large and complex societies ~\cite{naug2002role, Theis2015} due to their allowance for experimental manipulations ~\cite{baracchi2014socio, richardson2015beyond, ulrich2015distribution, stroeymeyt2018social}. However, there are significant limitations in groups with millions of individuals in underground colonies, as with leaf-cutter ants. Additionally, this group has an additional factor, the fungus garden, which is a fundamental part of these interactions. Mathematical models become especially interesting in these cases as they allow inferences for large colonies based on laboratory data.

Recent studies have associated Markovian processes with decision-making in ant colonies related to foraging ~\cite{baddeley2019optimal, friedman2021active, baltiansky2023emergent}. These models have been instrumental as they not only analyze the behavior itself and its consequences for the colony but also allow the association of which environmental elements trigger stimuli and the potential for these stimuli to affect the final response, which could lead to different strategies. This is the first work that applies this model to cleaning behaviors. Understanding how decision-making occurs in the presence of a pathogenic agent can help comprehend the mechanistic principles that unify the evolution of social immunity in eusocial organisms.

\section{Conclusion}
This study modeled the behavioral responses of ants challenged with an entomopathogenic fungus. Our results highlight how workers' interactions and past behaviors predict collective cleaning strategies. They reveal a strong link between cleaning and vulnerable limbs or pathogens, clarifying distinct behaviors and their transition times in different conditions. We discuss the importance of allogrooming, the potential implications for disease transmission, and possible changes to social contact networks. The conclusions highlight fixed decision-making principles in workers' behavior, highlighting the complex interaction between internal and external factors that influence social immunity strategies. This was the first study that applied Markov processes to understand ant grooming behavior, providing insights into mechanisms of social immunity and decision-making in eusocial organisms.
\section{Acknowledgments}
This publication has emanated from research conducted with the financial support of Brazilian Foundation CAPES, Brazilian Foundation CNPq, and Science Foundation Ireland under Grant 18/CRT/6049. The opinions, findings and conclusions or recommendations expressed in this material are those of the authors and do not necessarily reflect the views of the funding agencies.

\section{Declarations}

\textbf{Competing interests} The authors declare no conflicts of interest.

\textbf{Authors’ contributions} All authors conceived and designed the research. C.T. collected the data and provided insights into the discussion of results. G.R.P. and R.A.M. created the feature engineering methodology and analysed the data. G.R.P. led the writing of the manuscript. All authors contributed to the overall writing. 

\textbf{Funding} This work had the support from Science Foundation Ireland under grant number 18/CRT/6049; Brazilian Foundation CAPES process number 88887.716582/2022-00; Brazilian Foundation CNPq process number 141049/2020-0

\textbf{Availability of data and materials} All datasets and scripts are made available at \url{https://github.com/GabrielRPalma/UnderstandingLearningWithML}

\bibliographystyle{apalike}
\bibliography{ref}

\end{document}